\documentclass [aps,prl,twocolumn,showpacs] {revtex4} 

\usepackage[final]{graphics}
\usepackage{amssymb}
\usepackage{amsfonts}
\usepackage{epsfig}
 
\begin{document} 
 
%\preprint{APS/123-QED} 
 
\title{Monte Carlo Study of the  
Spin-1 Baxter-Wu Model} 

\author{M. L. M. Costa and J. A. Plascak}
\affiliation{Departamento de F\'\i sica, Instituto
de Ci\^encias Exatas, Universidade Federal de Minas Gerais, C. P. 702\\
30123-970, Belo Horizonte, MG - Brazil}

\email{marialuc@fisica.ufmg.br, pla@fisica.ufmg.br} 
 
\date{\today}% It is always \today, today, 
 
\begin{abstract} 
The two-dimensional spin-1 Baxter-Wu model is studied by using Monte Carlo
simulations. The standard single-spin-flip Metropolis algorithm is used
to generate the configurations from which the order parameter, specific heat
and magnetic susceptibility are measured. 
The finite-size scaling procedure
is employed in order to get the critical behavior. 
The extensive simulations shown that the critical exponents
are different from those of the spin-1/2 model suggesting that the spin-1
model is in a different universality class. 

\end{abstract} 
 
 
\maketitle 
 
\section{\label{sec:level1}Introduction} 

The  Baxter-Wu model is a system of spins defined on a two-dimensional  
triangular lattice with the classical spin variables $s_{i}$ taking only 
integer values. It was first introduced by Wood and Griffiths \cite{wood} as
a model which does not exhibit invariance by a global inversion of all
spins. The  system is described by the 
Hamiltonian 
\begin{eqnarray} 
{\cal{H}}=-J{\sum_{<ijk>}}s_{i}s_{j}s_{k}
%+D{\sum_{i}}{s_{i}}^{2} 
\label{hamil}, 
\end{eqnarray} 
where the  coupling constant 
$J$ is positive and the sum is over all triangles made up of 
nearest-neighbor sites on  the triangular lattice. 
For the spin-1/2 model, where  $s_{i}=\pm 1$, the exact solution 
obained by Baxter and Wu gives
$k_BT_c/J=2/\ln (1+\sqrt{2})$ and $\alpha = \nu ={{2}\over{3}}$\cite{ref1}. 
The system has also been studied with quenched impurities by Monte Carlo
\cite{mark1} and Monte Carlo renormalization group approaches \cite{mark2}.
Conformal invariance studies \cite{chico1,chico2} have shown that the pure spin-1/2
Baxter-Wu and the four-state Potts models have the same  operator content and are in
the same universality class. More recently, the short time critical dynamics has been
investigated through the relaxation of the order parameter at the critical temperature
by Monte Carlo simulations \cite{wag}. On the other hand,
for spin values greater or equal to one there are  neither exact solutions 
nor even much approximate approaches. 
It is the purpose of this work to study
the model above for the spin-1 case by using Monte Carlo simulations, where the variables 
$s_i$ take the values  $s_{i}=-1,0,1$.

Monte Carlo methods \cite{landau,barkema} form the largest and most important class of numerical 
methods used for solving statistical physics problems. The basic idea behind  
Monte Carlo simulation is to simulate the random thermal fluctuation of the 
system from state to state over the course of an experiment. Performing a 
high-precision finite-size scaling analysis using standard Monte Carlo techniques is  
very difficult due to constraints on the available computer resources. The 
introduction of histogram techniques to extract the maximum information 
from Monte Carlo simulation data at a single temperature enhances the potential 
resolution of Monte Carlo methods substantially \cite{alan1,alan2}.  
In this sense, we  apply the histogram techniques together with the  
Metropolis simulation algorithm in order to investigate the thermal behavior  
of the spin-1 Baxter-Wu model defined by Eq. (\ref{hamil}) by considering the specific heat, 
order parameter and magnetic  
susceptibility. Our main interest is to obtain, through a finite-size scaling
analysis,  the phase  
transition temperature as well as the critical exponents of the model.   
 
In the next section we  present the thermodynamic quantities and the details
of the simulations. In section III we discuss the results and in section IV we summarize our  
conclusions.

\section{Simulation background} 

The simulations have been carried out by using the single-spin-flip Metropolis 
algorithm. In the course of the simulations we considered triangular lattices
with linear dimensions $L \times L$ and fully periodic boundary conditions for
system sizes of length $18\le L \le 108$. Due to the fact that the system has,
in addition to the ferromagnetic phase (with all spins up), three different ferrimagnetic
phases with three different sublattices (one sublattice up and spins on the other
two sublattices down) the allowed values of $L$ are always a multiple of 3. In this way,
all ground states of the infinite lattice would fit on any finite lattice.
Following equilibration (which 
comprised $6 \times 10^4$ MCS) runs comprising up to $5 \times 10^6$ MCS
(Monte Carlo steps per spin) were performed. Histogram reweighting \cite{alan1,alan2}
and finite-size scaling techniques were used to precisely locate the second-order
phase transition. Regarding  the histograms, great care has been taken in order to
assure the reliabily of the extrapolated results for all lattice sizes.

The thermodynamic quantities we  measured in our simulations are
the order parameter, defined as the root mean square 
average of the magnetization of the three 
sublattices
\begin{equation}
m=\sqrt \frac{{m_{A}}^{2} + {m_{B}}^{2} + {m_{C}}^{2}}{3}~,
\label{mag}
\end{equation}
where $m_A$, $m_B$ and $m_C$ are the magnetizations per spin of the different
sublattices,
the order parameter susceptibility defined as
\begin{equation}
\chi  =  \beta  L^2 \left(\left<m^2\right> - \left< m \right >^2 \right)~,
\label{magsus}
\end{equation}
where $\beta =1/k_BT$ (with $k_B$ the Boltzmann constant and $\left<...\right>$ 
means an average over the generated Monte Carlo configurations),
%the quadrupole susceptibility
%
%
%\begin{equation}
%\chi _q =  \beta  L^2 \left(\left<Q^2\right> - \left< Q \right >^2\right)~,
%\label{quadsus}
%\end{equation}
%
%
%
%where
%
%\begin{equation}
%Q=\left< {{s_i^2}\over {L^2}}\right>~
%\label{quad}
%\end{equation}
%
%
%is the mean value of the square of the spin, and the specific heat
%
and the specific heat
\begin{equation}
C=\beta ^2 L^{-2}\left(\left<E^2\right>-\left<E\right>^2\right)~,
\label{spheat}
\end{equation}
where $\left<E\right>$ is the mean value of the energy.

According to  finite-size scaling theory the critical temperature scales
as
\begin{equation}
T_L=T_c+\lambda L^{-1/\nu}~,
\label{temp}
\end{equation}
where $\lambda$ is a constant, $T_c$ is the critical temperature of the
infinite system, and $T_L$ is the effective transition temperature for the lattice
of linear size $L$. This effective temperature can be given by
the position of the maximum of any of the following quantities:
the temperature derivative of $m$, $\ln m$ or $\ln m^2$, 
the order parameter susceptibility or the specific heat. The above temperatures
are given in units of $J/k_B$.
An independent estimate of $\nu$, however, can be made through the evaluation of the
maximum logarithmic
derivative of any power of the order parameter $m^n$
since one has
\begin{equation}
\left({{dU}\over{dT}}\right)_{max}=aL^{1/\nu}~,
\label{deriv}
\end{equation}
where $a$ is a constant and $U$ is either $\ln m$ or $\ln m^2$ (or, in general, $\ln m^n$).
In addition, the specific heat
and the magnetic susceptibility scale, at the transition temperature, as
\begin{equation}
C\propto L^{\alpha/\nu},~~~\chi\propto L^{\gamma/\nu}~,
\label{atTc}
\end{equation}
where $\alpha$ and $\gamma$ are the critical exponents of the specific heat and
susceptibility, respectively. From Eqs. (\ref{mag}-\ref{atTc}) one can obtain the
critical temperature and critical exponents of the model.
        
\section{Results} 
 
\begin{figure}[ht]
\includegraphics[clip,angle=0,width=7.7cm]{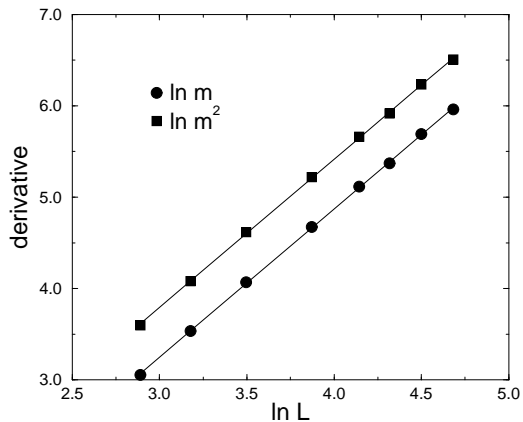}
\caption{\label{der} Logarithm of the maximum values of the derivatives of
$\ln m$ and $\ln m^2$ as a function of the logarithm of the size $L$.
The straight lines, with slopes corresponding to $\nu=0.617(3)$ in both cases, show the
asymptotic behavior of the fits. The errors are smaller than the symbol sizes.
}
\end{figure}

The independent evaluation of the critical exponent $\nu$, as obtained from
Eq. (\ref{deriv}) without any consideration of the critical temperature $T_c$, 
is shown in Fig. \ref{der} for the maximum derivative of the logarithm of $m$ and
$m^2$ (although other powers of $m$ can also be used). From both cases one
has $\nu=0.617(4)$, which is different from $\nu=0.666$ for the spin-1/2 model.

\begin{figure}[ht]
\includegraphics[clip,angle=0,width=7.7cm]{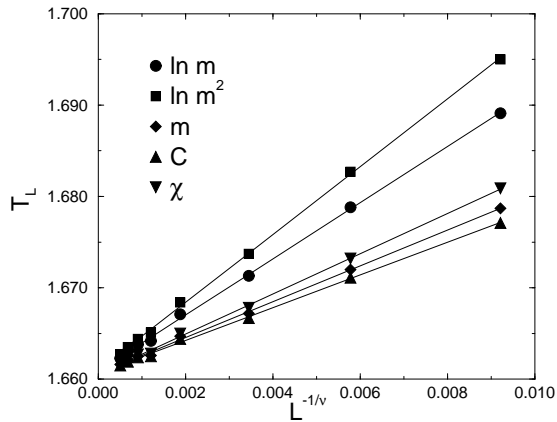}
\caption{\label{Tc} Size dependence of the effective critical temperatures 
(in units of $J/k_B$) estimated 
from several thermodynamic quantities. The lines are fits to Eq. (\ref{temp}) with 
$\nu=0.617$ obtained from Fig. \ref{der} and the intercepts are given in Table \ref{tab1}. 
The errors are smaller than the symbol sizes.
}
\end{figure}

With $\nu$ determined quite accurately we proceed to estimate the position of $T_c$.
As discussed in the previous section, the location of the maxima of the various 
thermodynamic derivatives, namely the maximum of the specific heat, susceptibility,
and the derivatives of $m$ and $\ln m$ and $\ln m^2$, provide estimates for the transition
temperature which scale with system size as Eq. (\ref{temp}). A plot of these estimates is
given in Fig. \ref{Tc}. The results from the linear fit are listed in Table \ref{tab1}.
One can note that they are indeed quite close to each other and a final estimate is 
$T_c=1.6607(3)$.

\begin{figure}[ht]
\includegraphics[clip,angle=0,width=7.7cm]{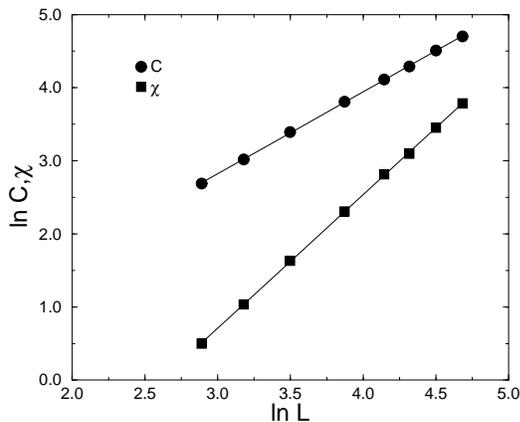}
\caption{\label{max} Logarithm of the maximum values of the specific heat $C$ 
and order parameter susceptibility $\chi$ as a function of the logarithm of $L$. 
The straight lines are
fits to Eqs. (\ref{atTc}) with $\alpha/\nu=1.121(6)$ and $\gamma/\nu=1.829(9)$.
The errors are smaller than the symbol sizes.
}
\end{figure}

The logarithm of the maximum value of the specific heat and order parameter susceptibility 
as a function of the logarithm of $L$ is shown in Fig. \ref{max}. From these data
one has $\alpha=0.692(8)$ and $\gamma=1.13(1)$.
%
%%%%%%%%%%%%%%%%%%%%%%%% fig.1 %%%%%%%%%%%%%%%%%%%%%%%%%%%%%%%%%%%%%%%%%%%%%%%%
%%%%%%%%%%%%%%%%%%%%%%%%%%%%%%%%%%%%%%%%%%%%%%%%%%%%%%%%%%%%%%%%%%%%%%%%%%%%%%%%%
%
%%%%%%%%%%%%%%%%%%%%%%%% fig.2 %%%%%%%%%%%%%%%%%%%%%%%%%%%%%%%%%%%%%%%%%%%%%%%%
%%%%%%%%%%%%%%%%%%%%%%%%%%%%%%%%%%%%%%%%%%%%%%%%%%%%%%%%%%%%%%%%%%%%%%%%%%%%%%%%%
% 
% 
\begin{table}[ht]\caption{Estimated critical temperatures from different thermodynamic
quantities according to the linear fit shown in Fig. \ref{Tc}.}
\begin{tabular}{cc}
\colrule  
Quantity &$    T_c$ \\
\colrule    
$C$             &  1.6607(1)    \\
$\chi$          &  1.6605(1)    \\
$\left({{dm}\over{dT}}\right)_{max}$         &  1.6606(1)    \\
$\left({{d\ln m}\over{dT}}\right)_{max}$     &  1.6609(1)    \\
$\left({{d \ln m^2}\over{dT}}\right)_{max}$  &  1.6610(2)    \\
\\
\colrule  
\end{tabular}\label{tab1}\end{table}

%
%%%%%%%%%%%%%%%%%%%%%%%% fig.3 %%%%%%%%%%%%%%%%%%%%%%%%%%%%%%%%%%%%%%%%%%%%%%%%
%%%%%%%%%%%%%%%%%%%%%%%%%%%%%%%%%%%%%%%%%%%%%%%%%%%%%%%%%%%%%%%%%%%%%%%%%%%%%%%%%
% 

\section{Conclusions} 

It is clear, from the quality of the above results, that a well defined second order phase
transition takes place in the model at $T_c=1.6607(3)$ with critical exponents
$\nu=0.617(3)$, $\alpha=0.692(6)$ and $\gamma=1.13(1)$,
which are indeed different from the spin-1/2 case, namely
$\nu=2/3$, $\alpha=2/3$ and $\gamma=7/6$. This means that this three spin interaction
model has exponents which depend on the spin value. It is worth saying that the
present model can also have an extra interaction with a crystal field of the form
$D{\sum_{i}}{s_{i}}^{2}$. This is a generalization in the direction of the so-called Blume-Capel
model \cite{blume}. What we have done here is  studied the special case $D=0$.
 However,
in the limit $D\rightarrow -\infty$ one recovers the spin-1/2 model. 
From the present results
we then expect that along the second-order transition line for different values of
$D$ one has a line with varying critical exponents. In addition, as we have shown, a 
second-order phase transition takes place at $D=0$ in contrast with the conjecture that
the spin-1 Baxter-Wu model is critical only in the limit $D\rightarrow -\infty$
\cite{kinzel}. Some preliminary results, agreeing with the picture of a line of
second-order phase transition with varying exponents 
and the presence of a multicritical point,
for the present system with crystal field
interaction, have already been obtained from conformal invariance with
finite-size  scaling theory and the mean field
renormalization group approach \cite{ze}. Work in this direction using Monte Carlo
simulations is now in progress.

\acknowledgments

We would like to thank R. Dickman
for fruitful discussions and a critical reading of the manuscript. 
Financial support from the Brazilian agencies
CNPq, CAPES, FAPEMIG and
CIAM-02 49.0101/03-8 (CNPq) are gratefully acknowledged.

\bibliography{apssamp}% Produces the bibliography via BibTeX. 

\end{document}